\documentclass[fleqn,usenatbib]{mnras}

\usepackage{newtxtext,newtxmath}

\usepackage[T1]{fontenc}
\usepackage{hyperref}

\DeclareRobustCommand{\VAN}[3]{#2}
\let\VANthebibliography\thebibliography
\def\thebibliography{\DeclareRobustCommand{\VAN}[3]{##3}\VANthebibliography}

\usepackage{graphicx}	
\usepackage{amsmath}	

\usepackage{booktabs}

\newcommand{\Msun}{\,{\rm M_\odot}}

\newcommand{\Mblack}{M_\bullet}
\newcommand{\dist}{\, \rm ckpc \, h^{-1}}

\title[Dynamics of Wandering IMBHs in the MW Galaxy]{Dynamics of Intermediate-Mass Black Holes Wandering in the Milky Way Galaxy Using the Illustris TNG50 Simulation}

\author[E. Weller et al.]{
Emma Jane Weller$^{1}$\thanks{emmaweller@college.harvard.edu},
Fabio Pacucci$^{1,2}\thanks{fabio.pacucci@cfa.harvard.edu}$,
Lars Hernquist$^{1}$ \&
Sownak Bose$^{1,3}$
\\
$^{1}$Center for Astrophysics $\vert$ Harvard \& Smithsonian, Cambridge, MA 02138, USA\\
$^{2}$Black Hole Initiative, Harvard University,
Cambridge, MA 02138, USA\\
$^{3}$Institute for Computational Cosmology, Durham University, South Road, DH1 3LE, Durham, UK
}

\date{\today}

\pubyear{2022}

\begin{document}
\label{firstpage}
\pagerange{\pageref{firstpage}--\pageref{lastpage}}
\maketitle

\begin{abstract}
The detection of Intermediate-Mass Black Holes (IMBHs) in dwarf galaxies is crucial to closing the gap in the wide mass distribution of black holes ($\sim 3 \Msun$ to $\sim 5 \times 10^{10} \Msun$). IMBHs originally located at the center of dwarfs that later collide with the Milky Way (MW) could be wandering, undetected, in our Galaxy. We used TNG50, the highest resolution run of the IllustrisTNG project, to study the kinematics and dynamics of star clusters, in the appropriate mass range, acting as IMBH proxies in a MW analog galaxy. We showed that $\sim 87\%$ of our studied IMBHs drift inward. The radial velocity of these sinking IMBHs has a median magnitude of $\sim 0.44 \, \mathrm{ckpc \, h^{-1} \, Gyr^{-1}}$ and no dependence on the black hole mass. The central $1 \dist$ has the highest number density of IMBHs in the galaxy. A physical toy model with linear drag forces was developed to explain the orbital circularization with time. These findings constrain the spatial distribution of IMBHs, suggesting that future searches should focus on the central regions of the Galaxy. Additionally, we found that the 3D velocity distribution of IMBHs with respect to the galactic center has a mean of $\sim 180 \, \mathrm{km \, s^{-1}}$ and larger variance with decreasing radius. Remarkably, the velocity distribution relative to the local gas shows significantly lower values, with a mean of $\sim 88 \, \mathrm{km \, s^{-1}}$. These results are instrumental for predicting the accretion and radiation properties of IMBHs, facilitating their detection with future surveys.
\end{abstract}




\section{Introduction} \label{Intro}

Black holes cover a wide range in mass and have been detected over a very large fraction of the known Universe. Stretching from $\sim 3\Msun$ \citep{Thompson_2019} to $\sim 5\times 10^{10}\Msun$ \citep{Dullo_2021}, they have been identified as close as $\sim 1000 \, \mathrm{pc}$ \citep{Oke_1977} and as far as $z=7.642$ \citep{Wang_2021}.

Two populations of black holes have been broadly investigated and located in specific environments. Stellar-mass black holes, with typical mass $\lesssim 10^3 \Msun$, are ubiquitous in the volume of galaxies. According to recent estimates (e.g., \citealt{Elbert_2018}), the Milky Way (MW) may host $\sim 10^8$ such black holes in its volume. Supermassive black holes (SMBHs), with masses $\gtrsim 10^6 \Msun$, are typically found at the center of massive galaxies. The MW hosts a SMBH of $\sim 4 \times 10^6 \Msun$ (e.g., \citealt{Ghez_2008, Genzel_2010}), and the near-horizon features of the central object in M87 were recently imaged by the EHT collaboration in a landmark result \citep{EHT_2019}.

Profound correlations exist between the mass of the central black hole and some properties of the host galaxy, such as its stellar mass and velocity dispersion \citep{Ferrarese_Merritt_2000, Gebhardt_2000}, suggesting a co-evolution between the massive central object and the host \citep{Kormendy_Ho_2013}. Extending these correlations to smaller black hole masses, it is reasonable to expect that black holes in an intermediate-mass range ($10^3 \Msun \lesssim \Mblack \lesssim 10^6 \Msun$), known as intermediate-mass black holes (IMBHs), should be contained in dwarf galaxies (see, e.g., the recent review by \citealt{Greene_2020_IMBH}). It is unclear whether these correlations hold down to very low masses, or break down instead (see, e.g, \citealt{Baldassare_2017_scaling, Pacucci_2018, Nguyen_2019, Baldassare_2020} for different perspectives on the topic). However, clear detections of IMBHs as light as $5 \times 10^4 \Msun$ are reported in dwarf galaxies \citep{Baldassare_2015}, and a sizable fraction of their nuclei are predicted to be active \citep{Pacucci_2021}, i.e., hosting accreting black holes.

The MW Galaxy has experienced numerous mergers during its cosmic history, which have possibly inserted IMBHs within its volume. Recent studies suggest that the MW has experienced $15 \pm 3$ mergers with galaxies with stellar masses $\gtrsim 4.5 \times 10^6 \Msun$ \citep{Kruijssen_2020}. These captured IMBHs could still be wandering inside the MW, with low-level accretion that makes them very challenging to detect. The investigation of these wandering and undetected IMBHs is fairly recent (see, e.g., \citealt{Bellovary_2010, Gonzales_2018, Greene_2021_wandering, Ricarte_2021b, Ricarte_2021a}). In particular, \cite{Ricarte_2021b} suggest that thousands of wandering black holes, also of supermassive size, might be found inside galaxy cluster halos.

In order to guide future observational efforts to detect IMBHs wandering in the MW, it is instrumental to understand their dynamics and kinematics. Where are IMBHs most likely located in the MW Galaxy, and what is their typical velocity? Furthermore, which forces act on them, and what is their typical sinking time? In this study, we use the TNG50 volume of the \href{https://www.tng-project.org/}{\color{cyan}IllustrisTNG project} \citep{Nelson_2019_Illustris} to provide some answers to these questions. 

It is important to remark that in this paper we focus only on IMBHs formed in dwarf galaxies and captured by the MW. Of course, other formation processes for IMBHs have been proposed, including: (i) runaway mergers in globular star clusters (e.g., \citealt{PZ_2002, Gurkan_2004, Gonzalez_2021, Shi_2021}), (ii) hyper-Eddington accretion onto stellar-mass black holes (e.g., \citealt{Ryu_2016}), (iii) direct collapse of hypermassive quasi-stars (e.g., \citealt{Volonteri_2010_IMBH, Schleicher_2013}), and (iv) supra-exponential accretion on seed black holes (e.g., \citealt{Alexander_Natarajan_2014, Natarajan_2021}).
Each of these channels would produce a different number density of IMBHs, as well as different spatial distributions, with likely very different dynamical properties. For example, IMBHs formed in-situ would probably have a lower typical velocity with respect to the local gas and stars. It is thus important to keep in mind that in this study we are focusing only on a specific formation channel for IMBHs.

We use the \cite{Planck_2020} cosmology as a reference.
In the following \S \ref{sec:simulation} we describe the simulation suite used, as well as our method to select IMBH proxies. In \S \ref{sec:location} we describe the location of IMBHs in the MW and the radial evolution of their orbits, while in \S \ref{sec:kinematics} we study their kinematics, regarding 3D and radial velocities, and dynamics. Finally, in \S \ref{sec:conclusions} we discuss the conclusions of our work.

\section{Simulation} \label{sec:simulation}
In this section we present the simulation suite and the methodology used to select IMBH proxies, as well as the MW analog galaxy.

\subsection{TNG50} \label{sec:TNG50}

IllustrisTNG \citep{Pillepich_methods, Pillepich_2018, Nelson_2018, Naiman_2018, Marinacci_2018, Springel_2018, Nelson_2019_Illustris} is a suite of cosmological simulations performed with the moving-mesh code \textsc{Arepo} \citep{Springel_2010}. Each simulation solves the evolutionary equations for dark matter, gas, stars, and SMBHs in a self-consistent fashion, from $z=127$ to $z=0$ \citep{Nelson_2019_Illustris}. For the purpose of the present study, we used TNG50-1 \citep{Nelson_2019, Pillepich_2019}, which is the simulation with the highest resolution. TNG50-1 has $2 \times 2160^3$ resolution elements, a volume of $35^3 \, \rm cMpc^3 \, h^{-3}$ (or $\sim 50^3 \, \rm cMpc^3$, hence the name), and a mean time of $\sim 0.138 \, \rm Gyr$ between snapshots. TNG50-1 also has three sub-boxes, defined as volumes of the simulation with data saved at a higher time cadence \citep{Nelson_2019}.

At $z=0$, the collisionless components (including star objects, which we use as IMBH proxies -- see \S \ref{sec:proxies}) have a gravitational softening length of $288 \, \rm pc$ \citep{Nelson_2019}. This is smaller than the scales on which we study black hole dynamics ($\gtrsim 1 \, \mathrm{kpc}$), so the relevant gravitational interactions are not affected. 

In TNG50 the baryon and dark matter particle masses are $8.5\times 10^4 \Msun$ and $4.5\times 10^5 \Msun$, respectively. Since our IMBH proxies have a mass range of $\sim (2.6 - 9.9) \times 10^4 \Msun$, this might suggest that the dynamical friction is possibly underestimated in certain situations, which then leads to a slight overestimation in the surviving IMBH population.
Nonetheless, the sinking timescales that we find are $\sim 10 \, \mathrm{Gyr}$, which is of the order that is expected for these objects under the influence of dynamical friction \citep{BT_1987}.

\subsection{Identifying Captured IMBH Proxies} \label{sec:proxies}

The black hole objects in the simulation mostly fall in the supermassive range -- the least massive black hole object in TNG50-1 at $z=0$ is $\sim 1.2 \times 10^6 \Msun$, which is larger than the mass range we aim to study. Instead, we used star (cluster) objects between $10^4$ and $10^6 \Msun$ as IMBH proxies. This is acceptable because we are only interested in gravitational effects on large scales. We used two methods to select IMBHs captured by the MW, which we describe in turn. 

In the first, we used a list of $z=0$ MW and M31 analog galaxies in the simulation \citep{Engler_2021, Pillepich_2021}. We identified a total of four MW analogs whose centers of mass fall within one of the three TNG50-1 sub-boxes at snapshot 99 ($z \approx 0$). We then searched the merger trees of these MW analogs for dwarf galaxies, defined as subhalos in the mass range $10^6 - 10^{10} \Msun$. We limited the search to more recent mergers, considering only dwarfs that appear after snapshot 60 ($\sim 7.4 \, \rm Gyr$) in the merger tree. We selected IMBH proxies from these dwarfs that end up in the corresponding MW analog (and within the sub-box) by $z=0$. We focused on IMBHs centrally-located in their host dwarfs, so we required the IMBHs to be less than $0.01 \, \rm ckpc$ from the dwarf center-of-mass (at the snapshot in which the dwarf appears in the MW analog merger tree). A total of four IMBHs, with masses in the range $\sim (3.4 - 8.5) \times 10^4 \Msun$, met all the above-mentioned conditions. We call this group Set 1. Extending the distance limit from the dwarf center-of-mass to $0.05 \, \rm ckpc$, we obtained 385 captured IMBH proxies, with masses between $\sim (3.0 - 9.1) \times 10^4 \Msun$. We call this group Set 2, and we use it to have a richer statistical population. All of the IMBH proxies in Sets 1 and 2 turned out to come from the same dwarf galaxy (Subhalo ID 456591 at snapshot 71, or $\sim 9.2 \, \rm Gyr$) and end up in the same MW analog (Subhalo ID 565089 at snapshot 99). This MW analog falls within sub-box 2, which contains $0.3 \%$ of the full simulation volume and has a mean time of $\sim 0.004 \, \rm Gyr$ between snapshots. We did not find a satisfactory explanation for why all of the suitable IMBHs came from the same dwarf, although it is important to note we only considered dwarfs that merged recently with just four MW analogs, so our collection was small to begin with.

To obtain a larger and more varied set of IMBHs for a larger statistical sample, we developed a second selection method. Rather than use the sub-boxes, we returned to the main simulation, but kept the same MW analog (Subhalo 565089) for consistency. We tested every 20th IMBH proxy in the galaxy at snapshot 99 ($z \approx 0$) and selected each one that, at the first snapshot after its birth, was not in the main progenitor of the MW analog. This resulted in 2,148 captured IMBH proxies, with masses between $\sim (2.6 - 9.9) \times 10^4 \Msun$, which we call Set 3. We find that the IMBHs in this set enter the MW over a period of 88 snapshots ($\sim 13.0 \, \rm Gyr$).

It is important to note that the mass range covered by our three sets is only $\sim 7.4 \%$ of the mass range we canonically assigned to IMBHs ($10^4$ to $10^6 \Msun$). This makes sense, as the largest star cluster object in the MW analog 565089 at $z=0$ is just $1.4 \times 10^5 \Msun$. This limited mass range is not ideal, but we believe it is better than using the black hole objects, which are far too large.

A summary of the characteristics of the three sets is given in Table \ref{tab:sets}. Two additional sets, both subsets of Set 3, are also included in the table, and described in \S \ref{Are IMBHs sinking?}.

Each star cluster acting as an IMBH proxy is tracked by its particle ID, which remains constant throughout the simulation. Single stars within a cluster cannot be tracked, because they do not possess individual identifiers in the Illustris TNG50 simulation. In order to track the position of IMBH proxies throughout the simulation, for each snapshot of interest an array of particle IDs of star cluster objects in the given MW analog was retrieved, along with arrays containing the particle properties of relevance. The properties of a given IMBH proxy were then found by locating the star cluster's particle ID against the full list of particle IDs.

\begin{table*}
\centering
\caption{Summary features of the five IMBHs sets employed.}
\begin{tabular}{ccc}
\hline
\textbf{Set \#}        & \textbf{IMBHs contained} & \textbf{Selection Criteria for IMBH Set}                                 \\ \hline
\multicolumn{1}{c|}{1} & 4                        & Distance $<0.01$ ckpc from dwarf center; obtained from sub-boxes \\ 
\multicolumn{1}{c|}{2} & 385                      & Distance $<0.05$ ckpc from dwarf center; obtained from sub-boxes \\ 
\multicolumn{1}{c|}{3} & 2,148                     & Captured IMBHs that originate outside MW analog; obtained from main simulation           \\ 
\multicolumn{1}{c|}{4} & 953                     & Subset of Set 3 that meets certain conditions described in \S \ref{Are IMBHs sinking?}            \\ 
\multicolumn{1}{c|}{5} & 825                     & Sinking IMBHs from Set 4           \\ \hline
\end{tabular}
\label{tab:sets}
\end{table*}

\subsection{MW Analog Galaxy} \label{sec:MW analog}

As explained in \S \ref{sec:proxies}, although there are many MW and M31-like galaxies in TNG50, only four lie in the sub-boxes, and our analysis uses just one of them -- namely Subhalo 565089. This was the only galaxy containing IMBHs that met the strict constraints for Sets 1 and 2. Thus, when we needed to loosen our selection criteria for Set 3 to obtain a larger number of IMBHs, focusing on IMBHs from Subhalo 565089 seemed like a natural choice. This also adds consistency across Sets 1-3, making results from the different sets more comparable. This is quite valuable, as we switch between the sets frequently throughout our study. 

In the following sections, we will refer to Subhalo 565089 as ``the MW analog” or simply ``the MW”. Additionally, the subhalo center-of-mass property is not available in the sub-boxes, hence we use the subhalo position for all relative motion calculations. Thus, ``MW center" refers to the position of Subhalo 565089.

\section{Location and radial evolution of Intermediate-Mass Black Holes} \label{sec:location}
In this section we describe the location of IMBHs in the MW, as well as their tendency to sink towards the center of the galaxy.

\subsection{Time Evolution of Radial Distance} \label{Time evolution}
\begin{figure}
	\includegraphics[width=\columnwidth]{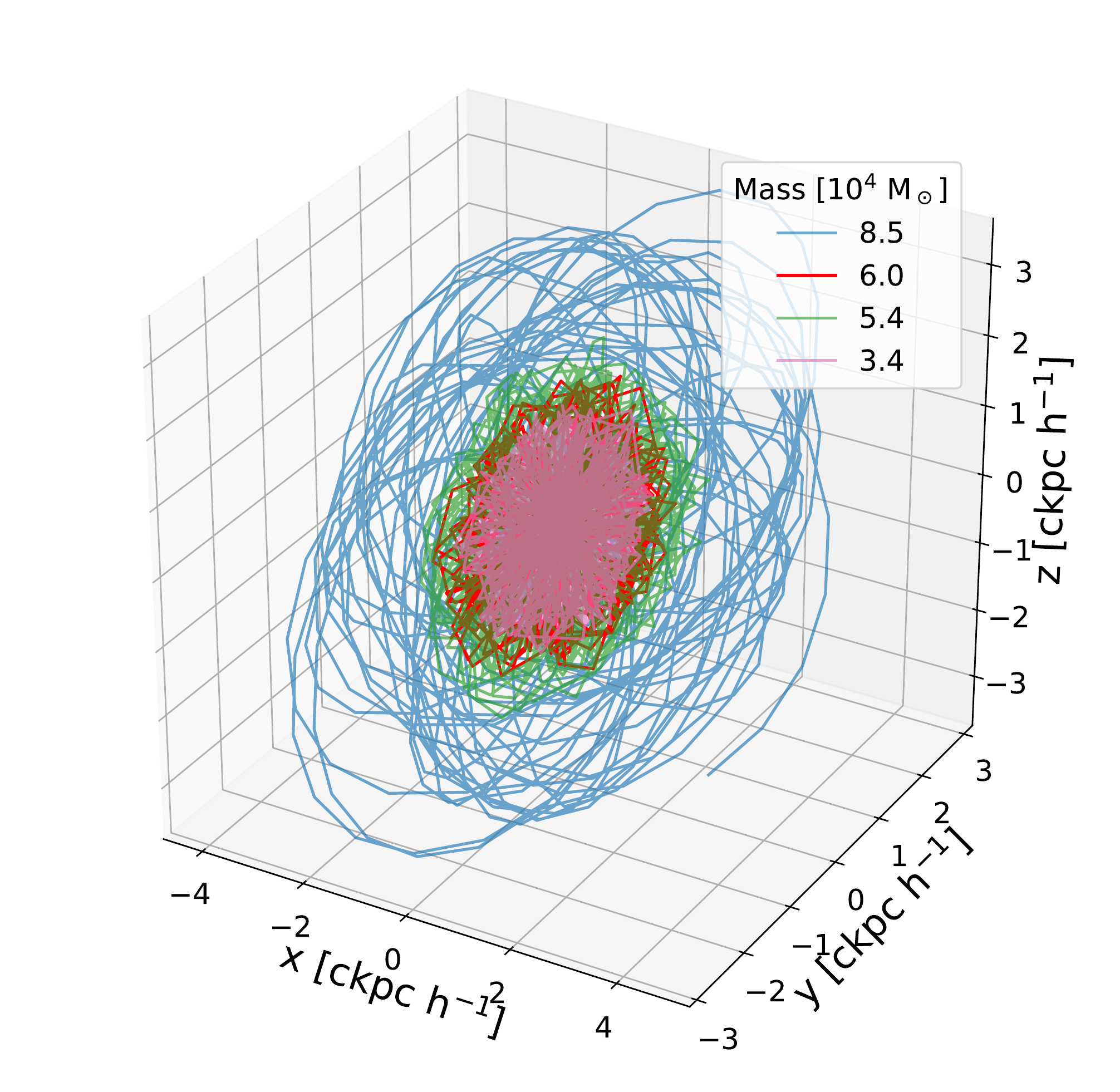}
    \caption{3D orbits of IMBHs in Set 1, for illustration purposes. These IMBHs originate in a dwarf galaxy, within $0.01 \, \rm ckpc$ of the center. The position of the IMBH is calculated with respect to the position of the MW center at every sub-box snapshot over a mean time period of $\sim 571$ snapshots ($\sim 4.42 \, \rm Gyr$).}
    \label{fig:Combined orbit}
\end{figure}

For each IMBH in Set 1, we calculated the $x$, $y$, and $z$ coordinates relative to the MW center at every sub-box snapshot from present time back to the time when the IMBH first entered the sub-box. These orbits are shown, for illustration purposes, in Fig. \ref{fig:Combined orbit}. 
These IMBHs have masses in the range $(3.4-8.5)\times 10^4\Msun$ and various orbital distances from the MW center.

\begin{figure}
	\includegraphics[width=\columnwidth]{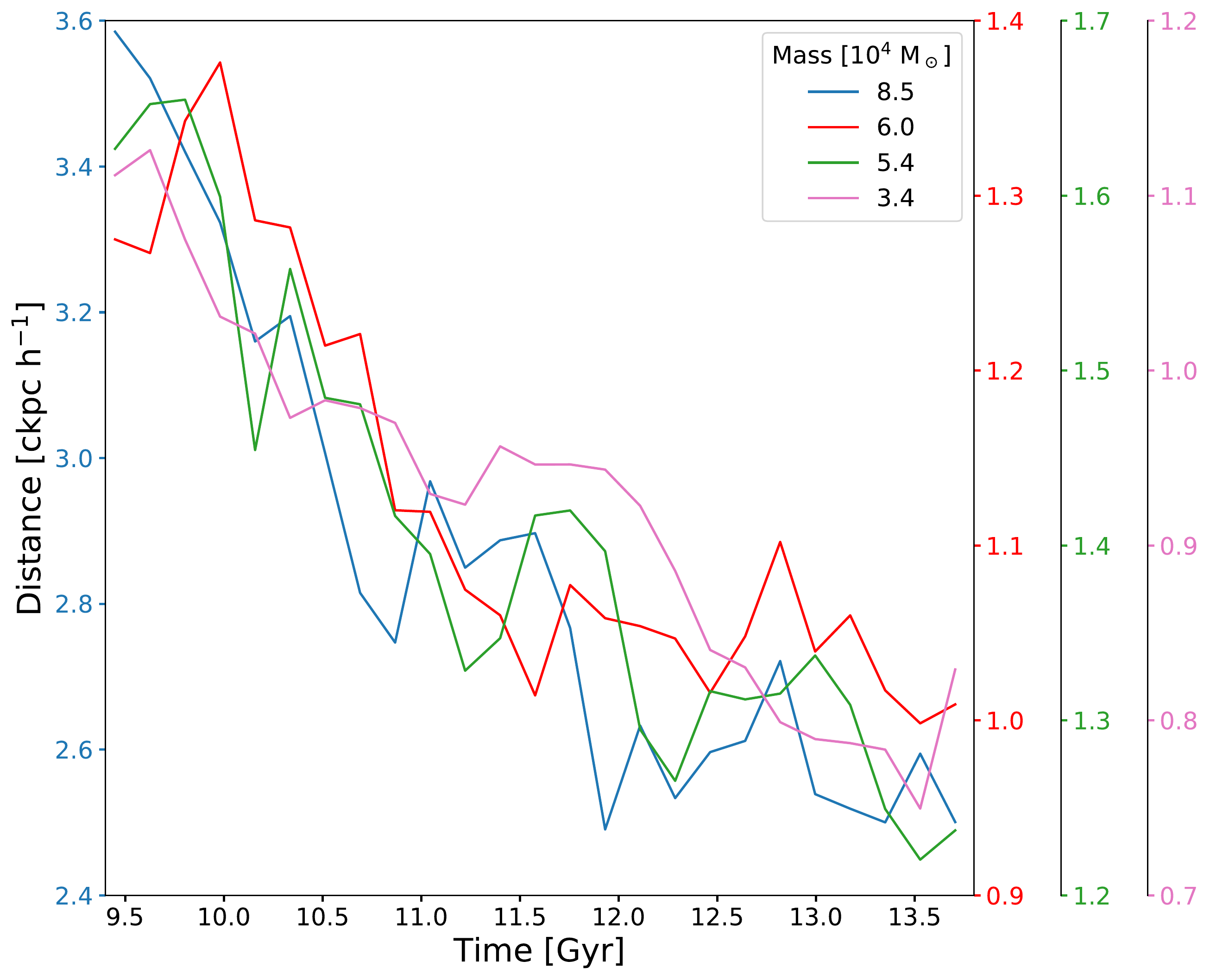}
    \caption{Time evolution of the radial distance of IMBHs in Set 1, with respect to the center of the MW. Distances are binned, with 25 bins and a mean bin size of $\sim 23$ sub-box snapshots ($\sim 0.177 \, \rm Gyr$). Four vertical axes are used to show overall trends over different distance ranges. For all four IMBHs, radial distance tends to decrease with increasing time.}
    \label{fig:Combined binned distance}
\end{figure}

\begin{figure}
	\includegraphics[width=\columnwidth]{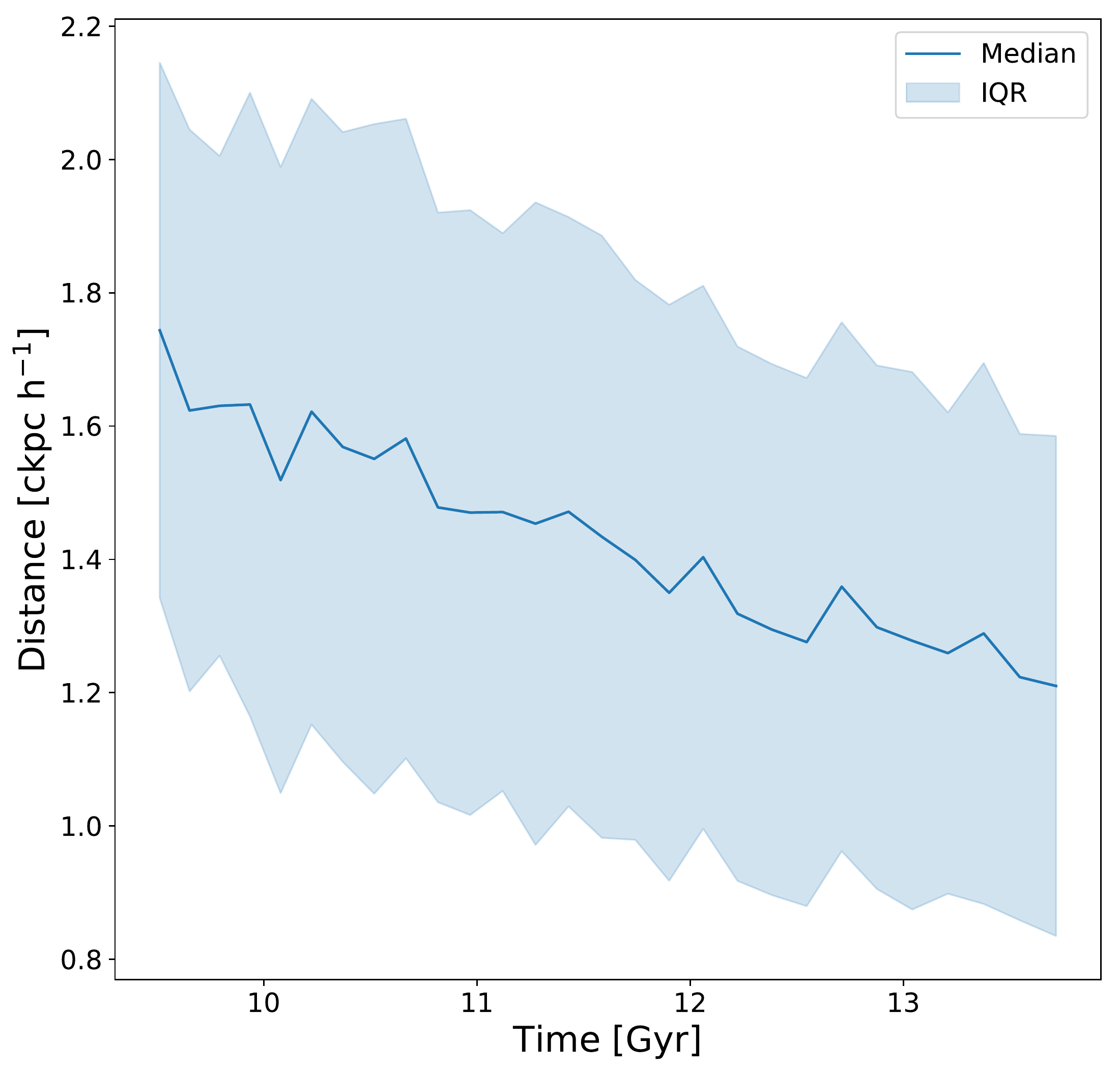}
    \caption{Distribution of radial distances from the center of the MW for IMBHs in Set 2, with respect to time. These IMBHs originate in a dwarf galaxy, within $0.05 \, \rm ckpc$ of the center. Both the median distance and the interquartile range (IQR) are shown. As time progresses, the radial distance tends to decrease.}
    \label{fig:Distance with IQR}
\end{figure}

We then used these data to calculate the distance from the MW center at each snapshot for each IMBH in Set 1. Binning the distances to reveal overall trends, we obtained the results shown in Fig. \ref{fig:Combined binned distance}.
We repeated the same process for Set 2, but rather than binning the distances, we only calculated the distance at every 20th sub-box snapshot, for computational simplicity. The results are shown in Fig. \ref{fig:Distance with IQR}, where the central solid line is the median, while the shaded range indicates dispersion via the interquartile range.

Figs. \ref{fig:Combined binned distance} and \ref{fig:Distance with IQR} indicate that the radial distances of IMBHs tend to decrease with increasing time. This fits with our expectation that captured IMBHs sink towards the galactic center, an effect that is explained by mass migration due to dynamical friction. Earlier works have already suggested that black holes, which are among the heaviest objects in galaxies, tend to cluster in the galactic center over time. For example,  \cite{Miralda_Escude_2000} suggested the presence of $\sim 25,000$ stellar mass black holes ($\Mblack \gtrsim 30 \Msun$) surrounding the central SMBH of the MW.

\subsection{Radial Velocity and Preferred Location of IMBHs} \label{Are IMBHs sinking?}

In the paragraphs above we showed that, on average, IMBHs are moving to smaller radii with time. To further study their long-term behavior, we calculated the secular radial velocity of a subset of the IMBHs in Set 3, i.e. our most extensive collection. In particular, we found the distance from the MW center for each IMBH at snapshots 81-84 ($\sim 10.8 - 11.3 \, \rm Gyr$) and 96-99 ($\sim 13.3 - 13.8 \, \rm Gyr$) of the main simulation. We chose this $\sim 3.0 \, \rm Gyr$ time period as a balance between avoiding small-scale variations while also refraining from assuming constant velocity on large timescales. 

We then calculated the best linear fit to this radial distance vs. time data for each IMBH. The slope gives the long-term radial velocity of the IMBH relative to the MW center, and the value of the line at $\sim 13.8 \, \rm Gyr$ gives the distance from the MW center at $z=0$.

To avoid any dynamic instability from IMBHs that just merged with the MW, we excluded any IMBHs that are not present in the MW for at least two snapshots prior to the studied time period (snapshots 79 and 80). We also excluded any IMBHs that are within $1 \dist$ of the MW center in snapshots 81-84, as well as any IMBHs whose estimated radial velocity changes by $1 \, \rm ckpc \, h^{-1} \, Gyr^{-1}$ or more when calculated using snapshots 81-84 and 88-91 (rather than 81-84 and 96-99). This helps ensure that we are not considering IMBHs that reach the central region of the MW before or near the beginning of the studied time period. These IMBHs would likely stay near the center for the entire time period, causing the radial velocity to be underestimated. Note that a few other IMBHs were excluded because, while they were present in the MW in snapshots 79 and 80, and still in the MW at snapshot 99, they were not classified as part of the MW in one or more of the other studied snapshots. There are 953 IMBHs that meet all of these requirements. We call this group Set 4 (see Table \ref{tab:sets}).

\begin{figure}
	\includegraphics[width=\columnwidth]{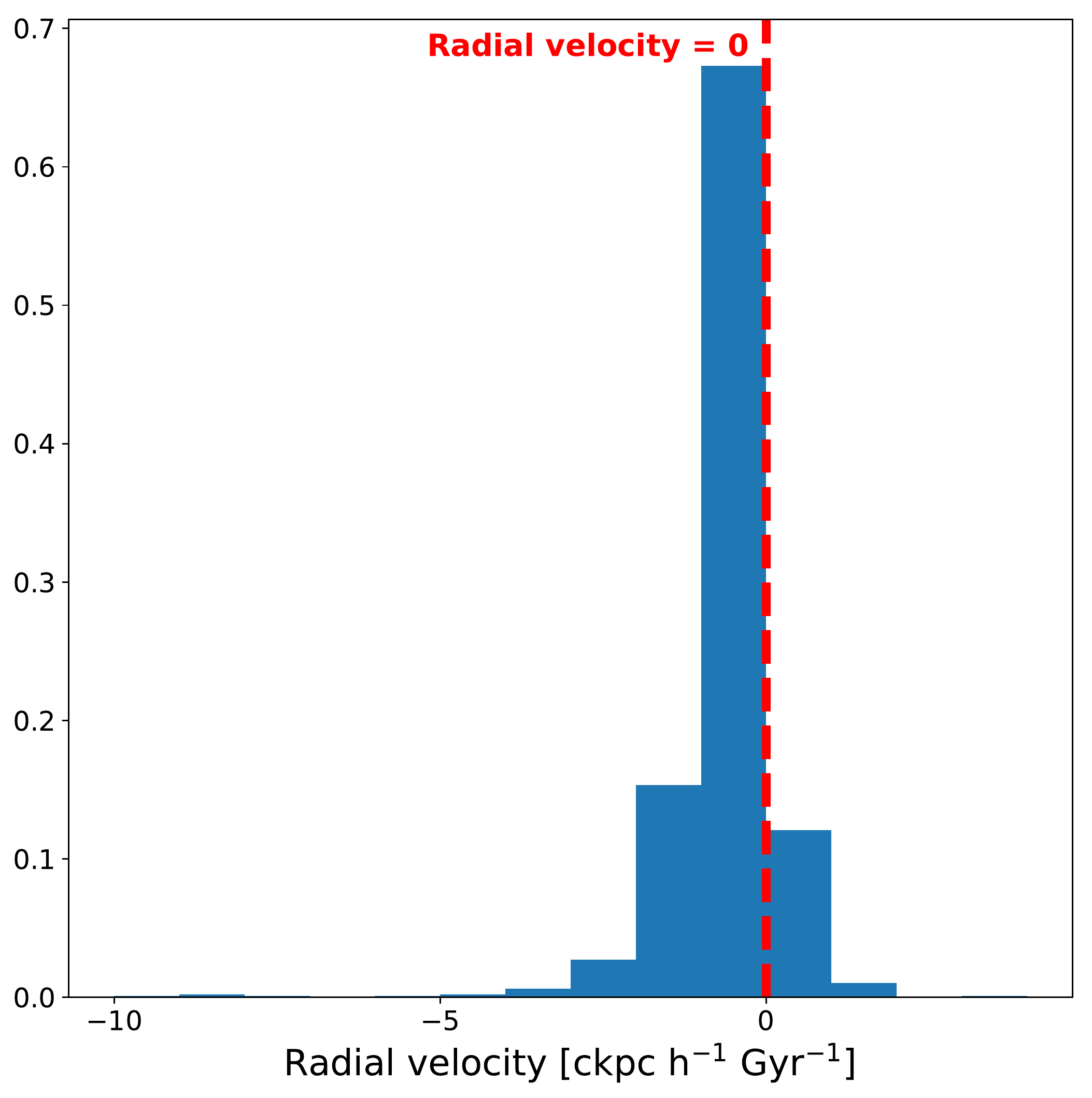}
    \caption{Long-term radial velocity distribution for IMBHs in Set 4 relative to the position of the MW center. These IMBHs originate outside the MW but are later captured, and they are subject to the restrictions described in \S \ref{Are IMBHs sinking?}. The radial velocities come from a linear fit to the radial distance vs. time data of each IMBH, calculated over eight snapshots spanning $\sim 3.0 \, \rm Gyr$. The vertical line indicates zero radial velocity. We find that $\sim 86.6 \%$ of the IMBHs have negative radial velocity, i.e., they are sinking.}
    \label{fig:Radial velocity histogram}
\end{figure}

Fig. \ref{fig:Radial velocity histogram} shows a histogram of radial velocities. The vertical values are normalized such that their integral is unity (from now on we will refer to this simply as a normalized histogram). We found that 825 IMBHs ($\sim 86.6 \%$ of Set 4) have negative radial velocity, with 640 ($\sim 67.2 \%$) between $-1$ and $0 \dist \, Gyr^{-1}$. We call the group of 825 sinking IMBHs Set 5 (see Table \ref{tab:sets}). The median radial velocity of this set is $\sim - 0.44 \, \mathrm{ckpc \, h^{-1} \, Gyr^{-1}}$. These results suggest that most captured IMBHs in the MW sink towards the galactic center.

\begin{figure}
	\includegraphics[width=\columnwidth]{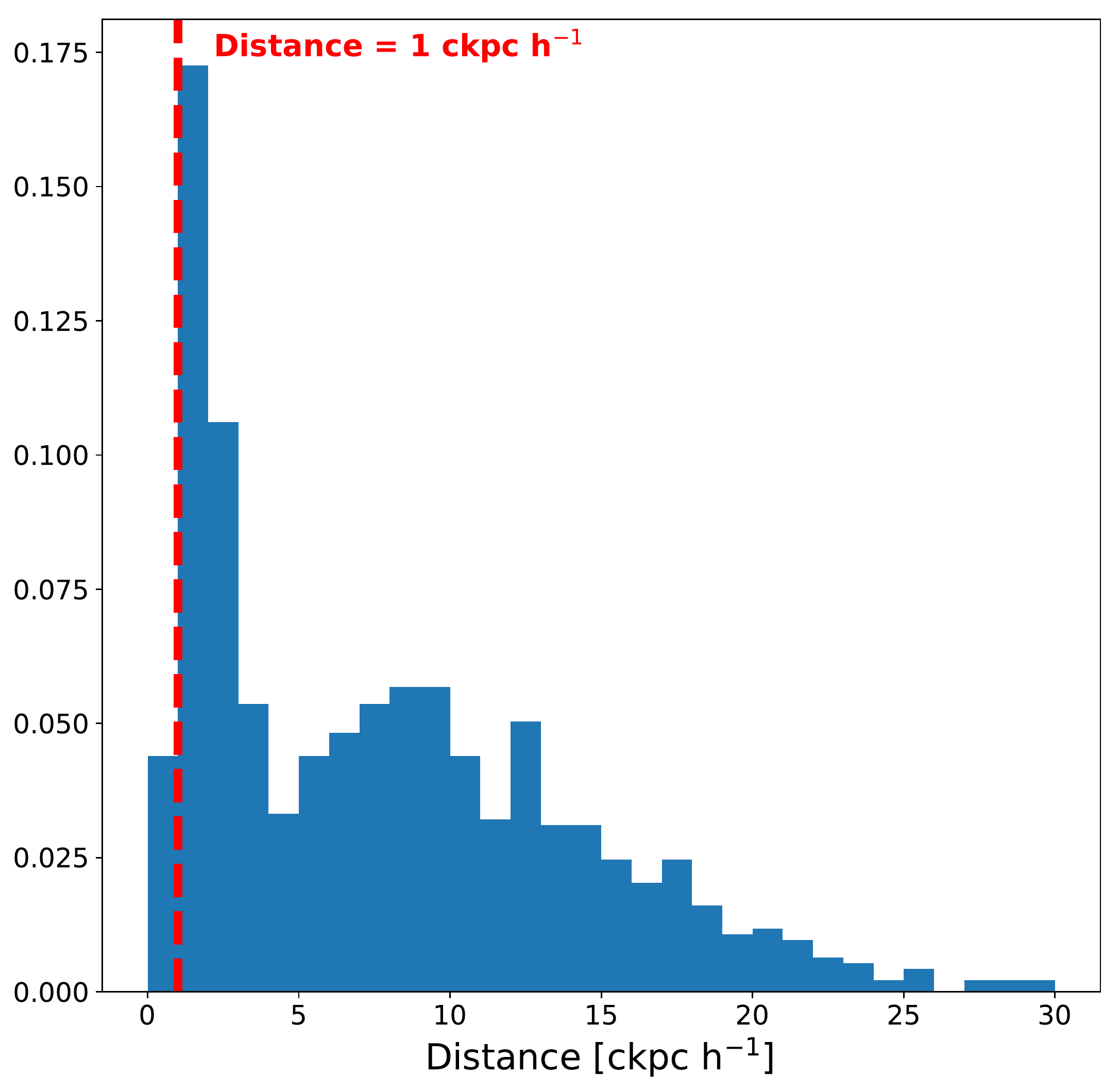}
    \caption{Distance distribution with respect to the MW center for IMBHs in Set 4. The distances are computed at $z=0$, using the linear fits to the radial distance vs. time data of each IMBH. The distance is cut at $30 \dist$ for a better visualization, with a vertical line indicating a distance of $1 \dist$. We find that $\sim 4.3 \%$ of the IMBHs fall within the innermost $1 \dist$, suggesting that the highest density of IMBHs lies in the central region of the Galaxy.}
    \label{fig:Distance at z=0 histogram}
\end{figure}

Fig. \ref{fig:Distance at z=0 histogram} shows a normalized histogram  of radial distances at $z=0$. We found that 41 IMBHs ($\sim 4.3 \%$ of Set 4) reach the central volume of the MW, defined as the innermost sphere with a radius of $1 \dist$. Recall that none of the IMBHs in Set 4 were in this region just $\sim 2.5 \, \rm Gyr$ earlier, at snapshot 84.

The resulting IMBH number density within the central $1 \dist$ is $\sim 1.6$ times the value found within the central $2 \dist$. As shown in the histogram, distance ranges past $2 \dist$ contain much smaller numbers of IMBHs. Thus, our simulations suggest that the central $\sim 1 \dist$ of the Galaxy holds the highest density of IMBHs.

To conclude, it is worth noting that, as our IMBHs in Sets 1 and 2 are chosen to be strongly bound to the host dwarf galaxy, none of them are ejected during the merging process with the MW analog. Interestingly, we computed that $\sim 7.5 \%$ of the IMBHs in Set 2 have a positive total energy with respect to the MW in snapshot 72 (the first snapshot after the merging event). Nonetheless, these IMBH proxies are still counted as part of the MW until the very end of the simulation, hence they are never practically ejected from the gravitational system.

\section{Kinematics and Dynamics of IMBHs in the MW} \label{sec:kinematics}
In this section we study the kinematics of IMBHs in the MW, with a particular focus on the distribution of their 3D velocities with respect to the center of the MW and with respect to the surrounding gas.

\subsection{Radial Distribution of 3D Velocities} \label{3D velocities}
\begin{figure}
	\includegraphics[width=\columnwidth]{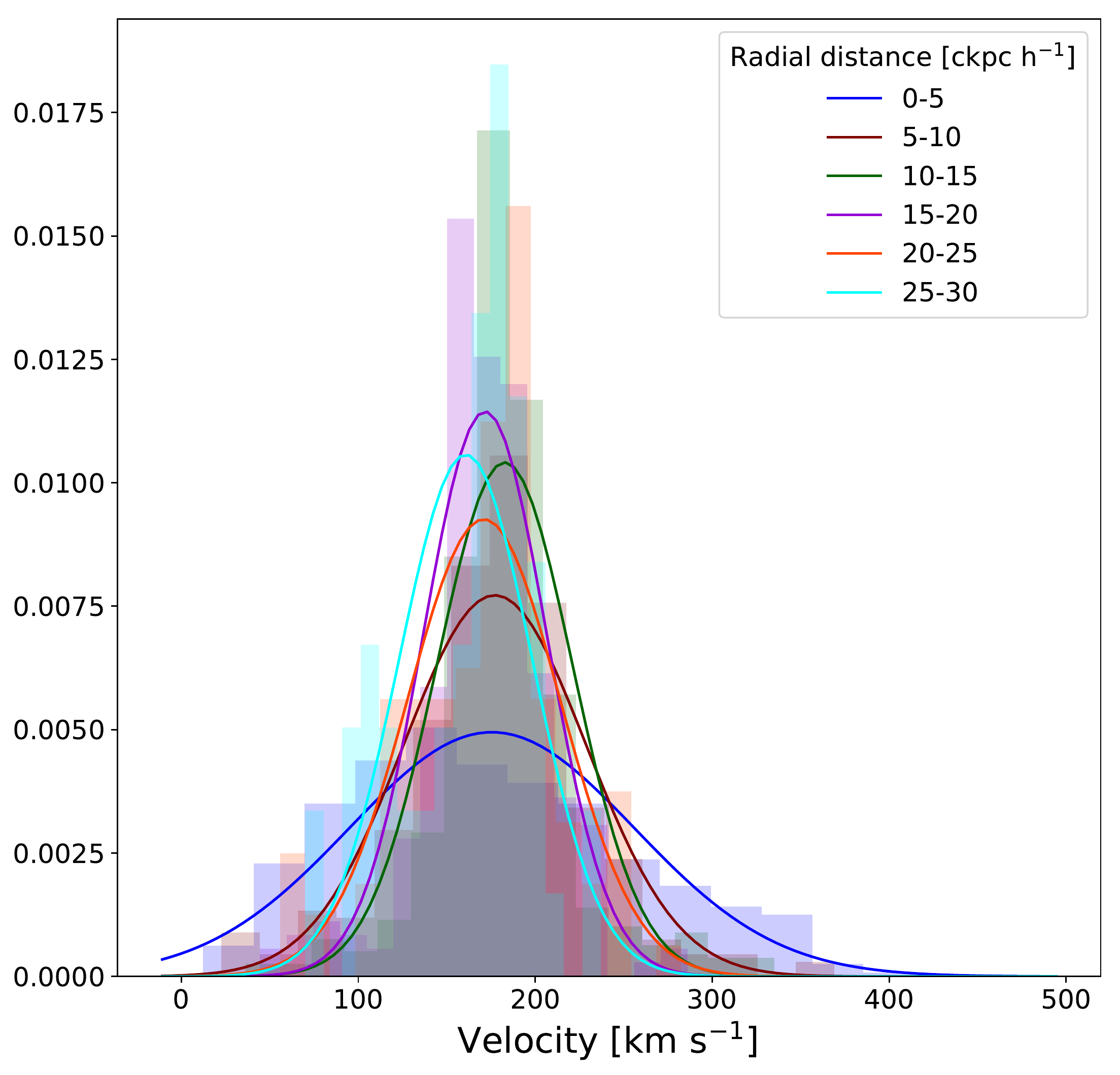}
    \caption{Distribution of 3D velocities for IMBHs in Set 3 with respect to the MW center, calculated at $z \approx 0$. These IMBHs originate outside the MW and are later captured. Radial distributions are divided into groups of $5 \dist$ each. Gaussian fits to each velocity distribution are provided.}
    \label{fig:Velocity magnitude histogram}
\end{figure}

To study the radial velocity distribution, we used Set 3, our most extensive IMBH catalog, and worked in snapshot 99 ($z \approx 0$). For each IMBH, we calculated the distance from the MW center and the magnitude of the velocity vector relative to the MW center. We then sorted the velocity magnitudes into groups by their corresponding radial distances and found a Gaussian fit for the velocity distribution of each group. Fig. \ref{fig:Velocity magnitude histogram} shows normalized histograms (with 16 bins each) and Gaussian fits for groups with corresponding radial distances up to $30 \dist$. Table \ref{tab:velocity_distributions} gives the statistical descriptors of these Gaussian fits.

\begin{figure}
	\includegraphics[width=\columnwidth]{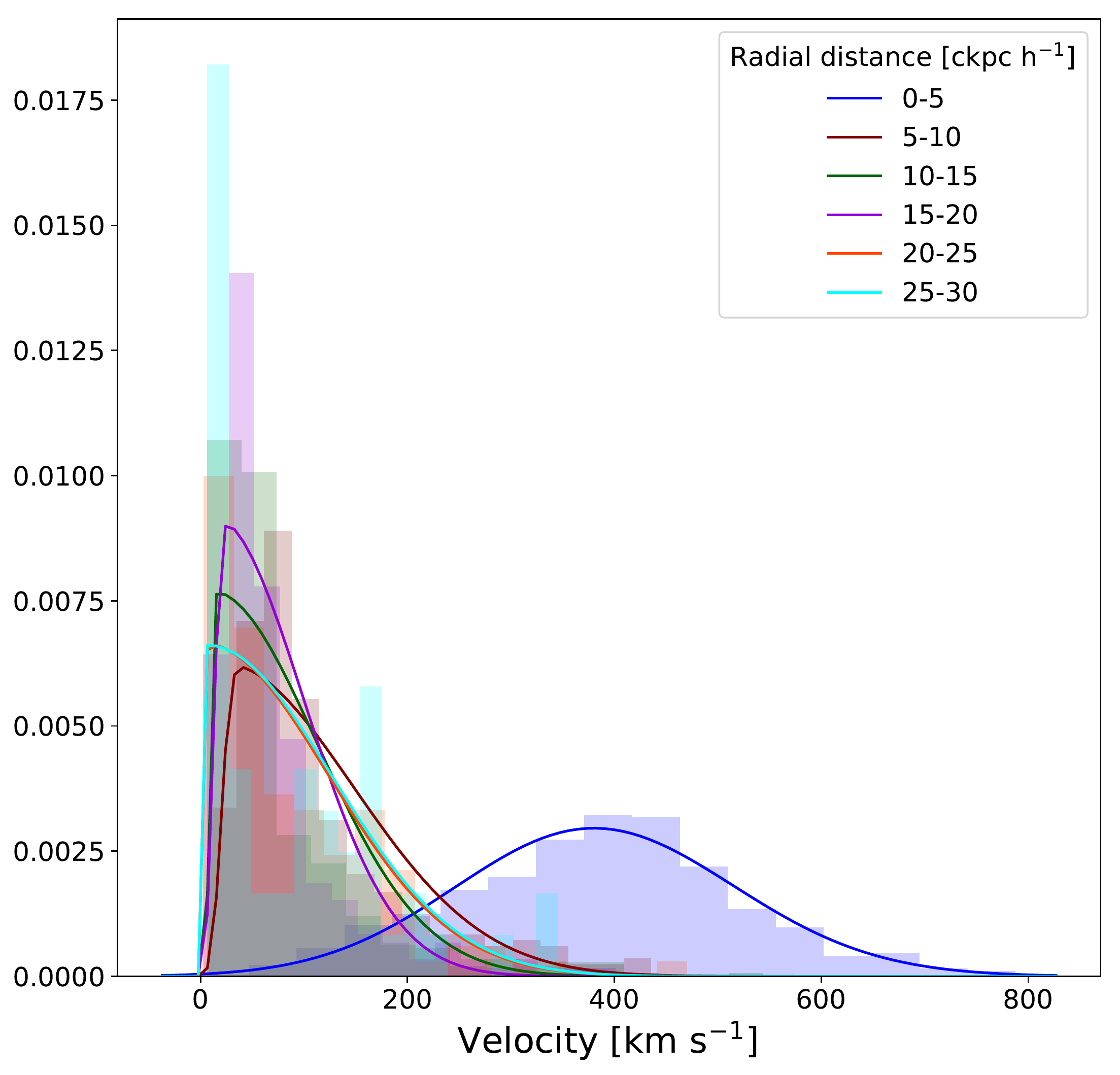}
    \caption{As in Fig. \ref{fig:Velocity magnitude histogram}, but with velocities calculated with respect to the local gas frame. Skewed Gaussian fits are also provided. This distribution is useful for calculating the accretion rates onto IMBHs wandering in the MW.}
    \label{fig:Gas velocity skewed fit}
\end{figure}

Next, for every IMBH, we calculated the velocity vector relative to each of the four closest gas particles. We then calculated the mean of the magnitudes of the four velocity vectors for each IMBH and, as before, sorted the mean velocity magnitudes into groups based on radial distance from the MW center. Fig. \ref{fig:Gas velocity skewed fit} shows normalized histograms (with 16 bins) for each group up to $30 \dist$. Also included are skewed normal fits for the velocity distribution of each group. Table \ref{tab:velocity_distributions} gives the statistical descriptors of these skewed fits. In implementing the skewed normal distribution, we used the definition provided by \cite{Azzalini_Capitanio_2009}.

We recognize that the histograms in Fig. \ref{fig:Velocity magnitude histogram} are not necessarily Gaussian, but we use this distribution as the simplest one to obtain some statistical information on the mean and variance of the histograms. With a Gaussian fit we intend to provide a simple description of the data, not to infer any physical interpretation on its origin. The Gaussian fits in Fig. \ref{fig:Velocity magnitude histogram} are also used to highlight the strong contrast with the distributions of velocities relative to the local gas (Fig. \ref{fig:Gas velocity skewed fit}), which are, instead, strongly non-Gaussian (except the radial bin $0-5 \dist$) and fitted with a skewed function.

Fig. \ref{fig:Velocity magnitude histogram} shows that typical velocities of IMBHs with respect to the center of the MW are $\sim 180 \, \mathrm{km \, s^{-1}}$, with larger variance as the radial distance decreases. In particular, we note that the variance is a factor $\sim 2$ larger in the innermost radial bin than in the outermost bin. This may be due to a larger number of IMBHs in the inner bins, but a higher stellar velocity variance with decreasing galactocentric distance is also observed and predicted by models \citep{Battaglia_2005, Dehnen_2006, Brown_2010}.

Fig. \ref{fig:Gas velocity skewed fit} shows that velocity magnitudes relative to the local gas are significantly lower, typically $\sim 88 \, \mathrm{km \, s^{-1}}$ (for radial distances $\gtrsim 5 \, \mathrm{ckpc \, h^{-1}}$), and the distribution is highly skewed towards high values.
The distribution of IMBH velocities with respect to the local gas was obtained because it is relevant for the calculation of accretion rates onto wandering IMBHs in the MW. In fact, classical Bondi accretion calculations \citep{Bondi_1952} rely on the velocity of the wandering object with respect to the local gas frame. More details on this issue are discussed in \S \ref{sec:conclusions}.

\begin{figure}
	\includegraphics[width=\columnwidth]{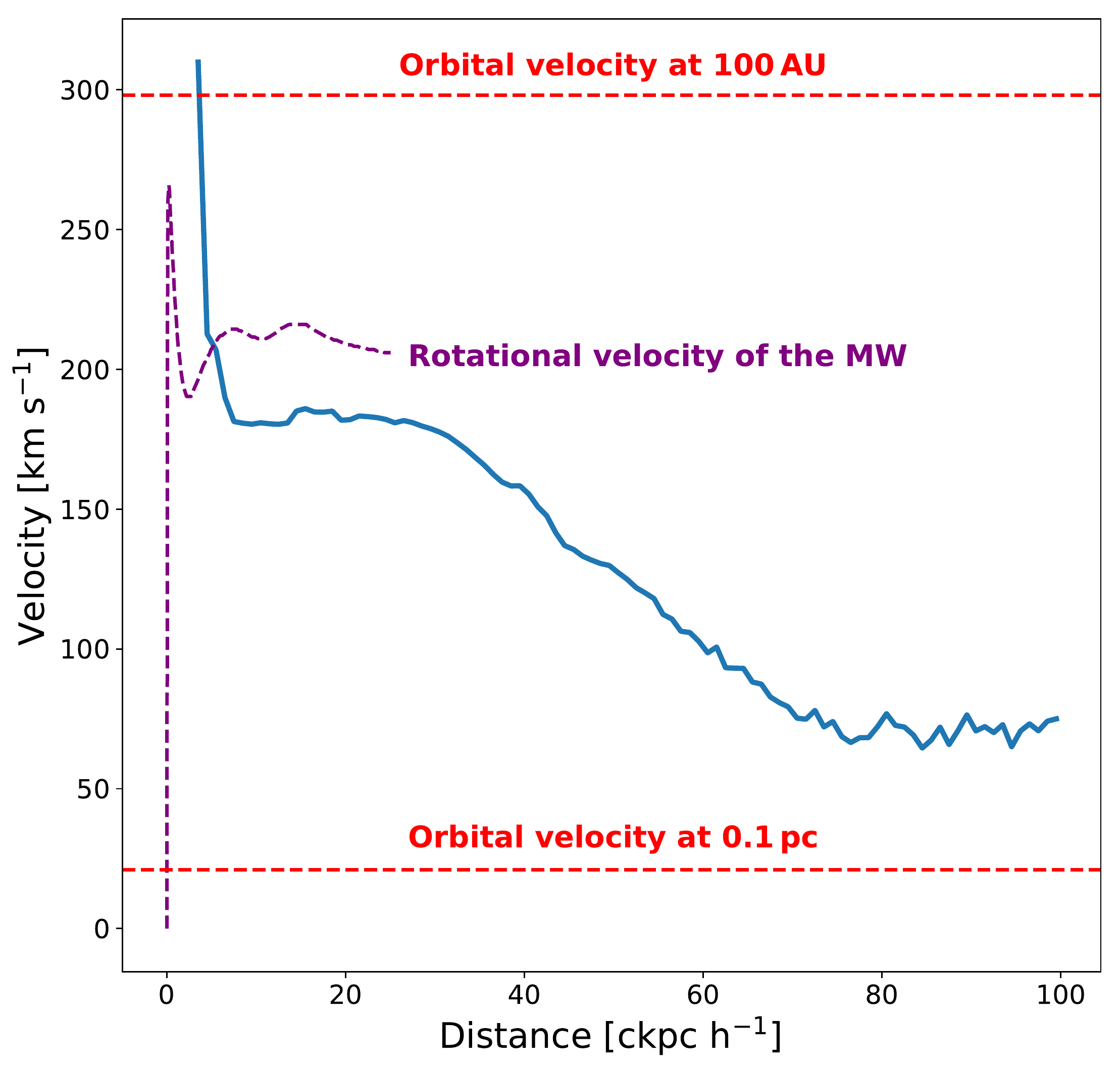}
    \caption{Median gas velocity magnitude (relative to MW center) vs. radial distance. The two horizontal, dashed lines represent the orbital velocity around a $10^8 \Msun$ SMBH, at $100 \, \rm AU$ and at $0.1 \, \rm pc$, for reference. Also shown is the rotational velocity of the MW up to $25 \, \rm kpc$ \citep{Li_2016}.}
    \label{fig:Median gas velocity}
\end{figure}

To explain the high velocity of the $0-5 \dist$ range in Fig. \ref{fig:Gas velocity skewed fit}, we found the radial distance and velocity magnitude relative to the MW center for each gas particle in the MW. We then sorted the velocity vs. distance data into $1 \dist$ bins and found the median velocity of each group. The results, shown in Fig. \ref{fig:Median gas velocity}, suggest that the gas velocity in the central $5 \dist$ of the MW is significantly high, and explained by the presence of a $\sim 9 \times 10^7 \Msun$ SMBH at the center of the MW analog. Fig. \ref{fig:Median gas velocity} also shows the orbital velocity around a typical $\sim 10^8 \Msun$ SMBH at radial distances of $100 \, \rm AU$ and $0.1 \, \rm pc$, as well as the rotational velocity of stars in the MW up to $25$ kpc from \cite{Li_2016}.

\begin{table*}
\centering
\caption{Statistical descriptors of the velocity distribution of IMBHs with respect to the center of the MW and to the local gas. The data relative to the MW center are fitted with a Gaussian probability distribution function. The data relative to the local gas are fitted with a skewed probability distribution. All values are rounded to two decimals.}
\label{tab:velocity_distributions}
\begin{tabular}{cccccc}
\hline
 & \multicolumn{2}{c}{\textbf{MW center}} & \multicolumn{3}{c}{\textbf{Local gas}} \\
\cmidrule(lr){2-3} \cmidrule(lr){4-6}
\textbf{Distance range {[}$\mathrm{ckpc \, h^{-1}}${]}} & \textbf{Mean {[}$\mathrm{km \, s^{-1}}${]}} & \textbf{Std. deviation {[}$\mathrm{km \, s^{-1}}${]}} & \textbf{Mean {[}$\mathrm{km \, s^{-1}}${]}} & \textbf{Std. deviation {[}$\mathrm{km \, s^{-1}}${]}} & \textbf{Skewness parameter} \\ \hline

$\mathbf{0-5}$                               & $175.46$                                       & $80.64$ & $384.25$                                       & $135.11$       & $0.69$                                                       \\ 
$\mathbf{5-10}$                              & $177.15$                                       & $51.66$ & $114.22$                                       & $85.70$        & $18.30$                                                      \\ 
$\mathbf{10-15}$                             & $182.73$                                        & $38.31$ & $77.42$                                        & $77.47$        & $38.22$                                                      \\ 
$\mathbf{15-20}$                             & $171.58$                                        & $34.86$ & $71.80$                                        & $63.34$        & $16.76$                                                      \\ 
$\mathbf{20-25}$                                      & $170.90$                                        & $43.07$ & $90.17$                                        & $82.88$        & $75.03$                                                      \\ 
$\mathbf{25-30}$                                      & $160.69$                                        & $37.74$ & $92.11$                                        & $84.62$        & $6839254.74$                                                      \\ \hline
\end{tabular}
\end{table*}

\subsection{Orbital Circularization} \label{Sinking velocities}

\begin{figure}
	\includegraphics[width=\columnwidth]{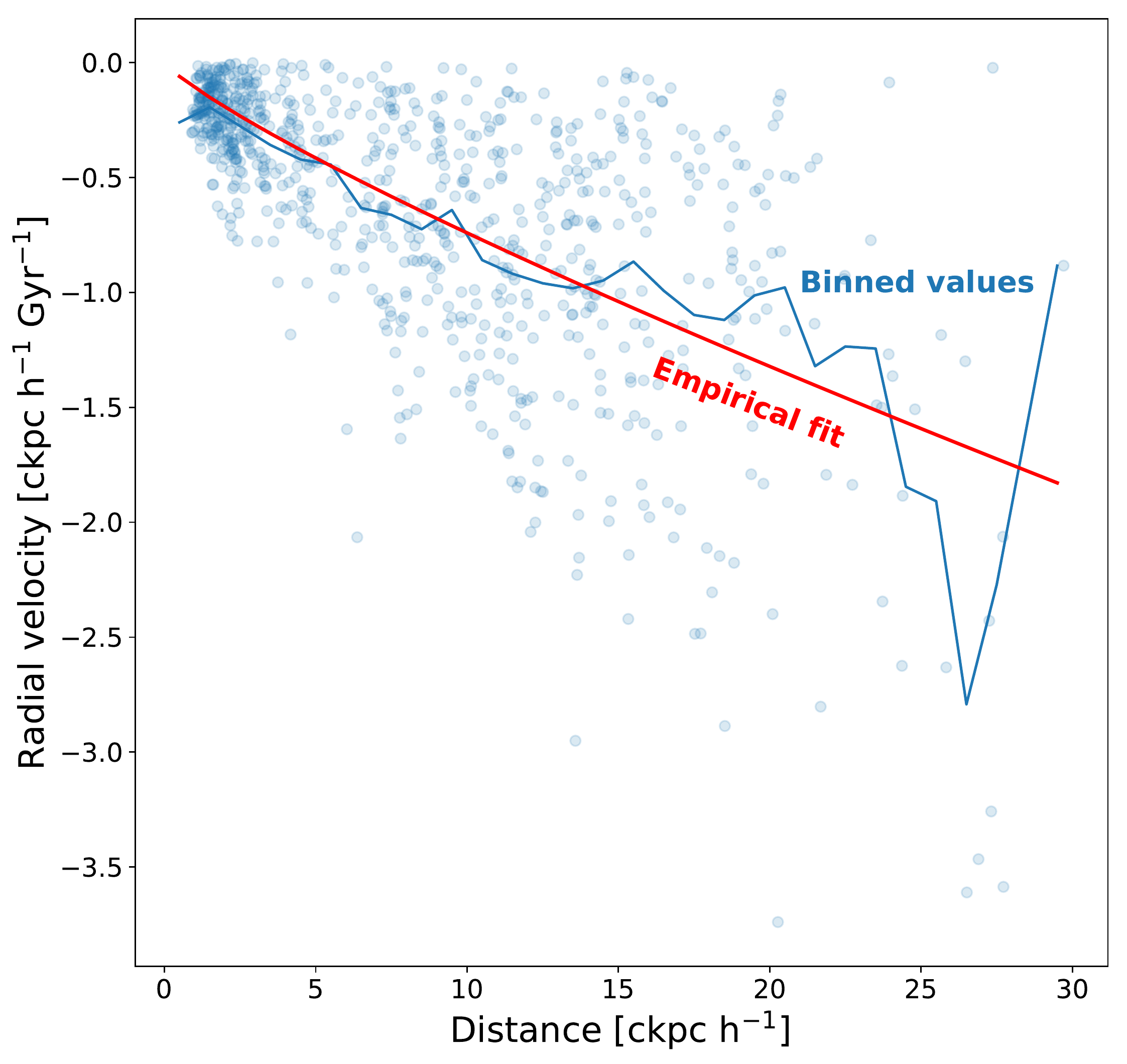}
    \caption{Sinking velocity vs. radial distance relative to the MW center for IMBHs in Set 5 with radial distances less than $30 \dist$. These are captured IMBHs with negative radial velocity, subject to certain conditions described in \S \ref{Are IMBHs sinking?}. The data is shown as a scatter plot, with binned values and the empirical fit also provided. The fact that sinking velocities tend to get smaller with decreasing distance from the galactic center indicates that IMBH orbits circularize as they sink farther into the MW.}
    \label{fig:Sinking velocity vs. distance fit}
\end{figure}

We now turn our attention to sinking IMBHs. For each IMBH in Set 5, we found the value of the best fit line from \S \ref{Are IMBHs sinking?} at $\sim 12.3 \, \rm Gyr$ (the mean of the eight times at which radial distance was measured). This gave us an estimate for the distance from the MW center. We then considered sinking (radial) velocity vs. radial distance for the 808 sinking IMBHs with radial distances less than $30 \dist$. Fig. \ref{fig:Sinking velocity vs. distance fit} shows a scatter plot of the data, as well as the mean sinking velocity over $1 \dist$ bins (darker blue line). We found the binned data to be well-fitted by an empirical curve of the form:
\begin{equation} 
        V = k R^p \, ,
\label{eq:V_eq}
\end{equation}
where $V$ and $R$ are the sinking velocity in $\rm ckpc \, h^{-1} \, Gyr^{-1}$ and radial distance in $\dist$, respectively, both functions of the simulation time $t$ in $\rm Gyr$. We found $k \approx -0.10846 \, \rm Gyr^{-1}$ and $p \approx 0.83472$. The results of the regression are added to Fig. \ref{fig:Sinking velocity vs. distance fit} as a red line.

\begin{figure}
	\includegraphics[width=\columnwidth]{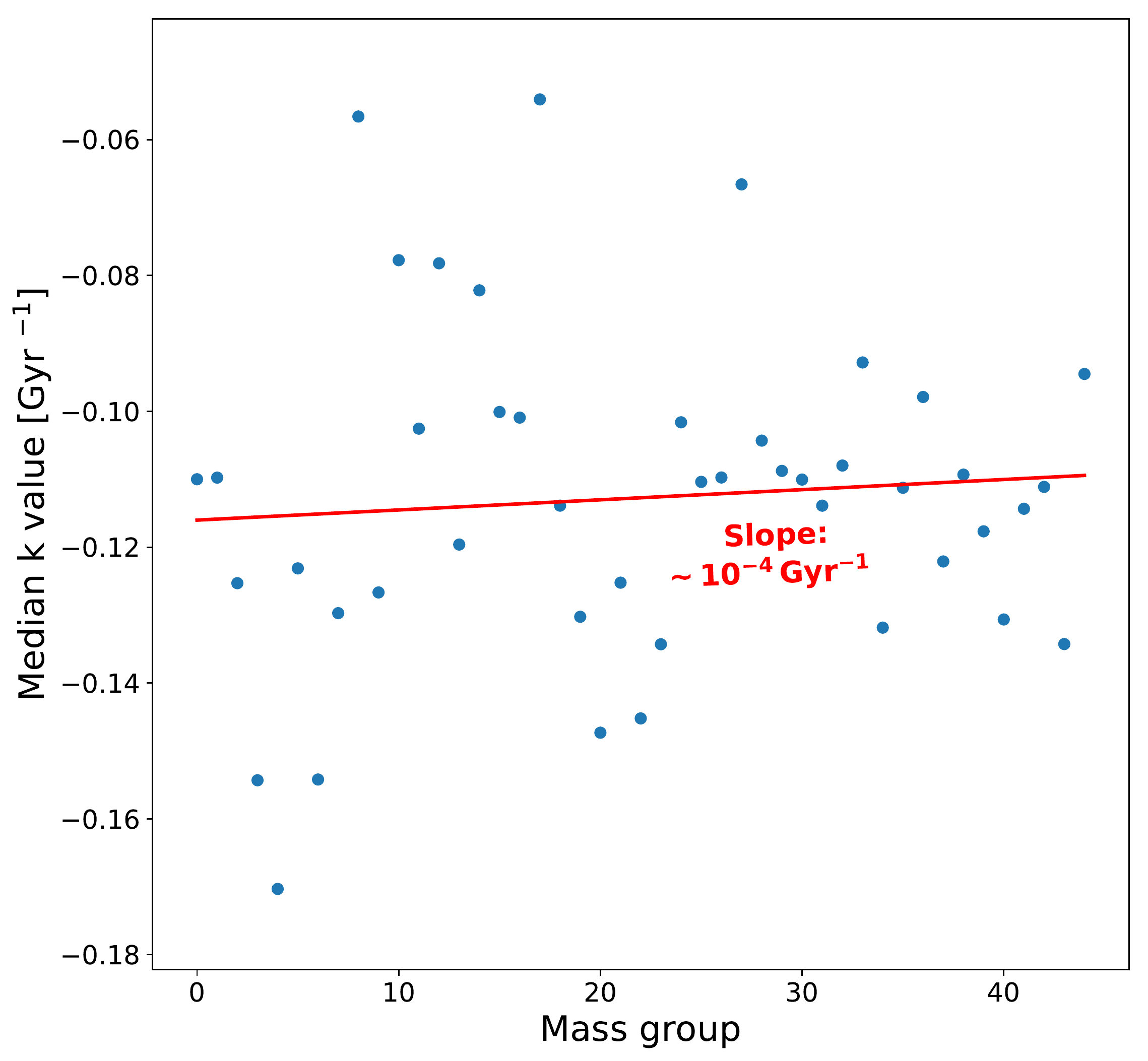}
    \caption{Median $k$ value vs. group number. Here, $k$ is the constant of proportionality in the empirically derived relation $V = k R^p$ between sinking velocity and radial distance. IMBHs in Set 5 with radial distances less than $30 \dist$ are divided by mass into 45 groups, where the group number increases with increasing mass. The best fit line to this data is shown, and the very small slope indicates that sinking velocity does not depend on black hole mass.}
    \label{fig:k vs. mass group}
\end{figure}

The fact that the magnitude of the radial velocity $V(t)$ declines as an IMBH approaches the center matches our expectation that the orbits of the IMBHs circularize as they sink farther into the MW (for a semi-analytical treatment of this effect, see, e.g., \citealt{Bonetti_2020}).
Solving the ordinary differential equation in Eq. \ref{eq:V_eq}, with integration constant $R(t=0) = R_0$, we find:
\begin{equation} 
        R(t) = \left[ R_0^{(1-p)} + k \, t (1-p) \right]^\frac{1}{1-p} \, .
\label{eq:R_empirical}
\end{equation}
We divided this IMBH dataset into 45 mass bins and calculated the median value of $k = V/R^p$ (where we keep $p \approx 0.83472$) for each group. We plotted the median $k$ values vs. group number, as shown in Fig. \ref{fig:k vs. mass group}, and obtained a best fit line with a slope of $\sim 1.4951 \times 10^{-4} \, \rm Gyr^{-1}$, indicating virtually no dependence on mass. In other words, the sinking velocity does not depend on the mass of the IMBH.
This result is interesting because one would naively expect heavier black holes to sink faster. In reality, IMBHs sink inefficiently towards the center, especially when wandering inside clumpier galaxies. For example, \cite{Ma_2021} have recently suggested that in high-$z$ clumpy galaxies, black holes less massive than $10^8 \Msun$ (a limit that includes our entire mass range) migrate inefficiently towards the center, with velocity magnitudes changing erratically and a much slower deceleration.

The empirical model in Eq. \ref{eq:V_eq} is useful to determine general trends in the data, but it does not provide any information regarding the forces acting to reduce the radial velocity as the IMBH sinks. Below, we develop a toy model to study the circularization effect and gain useful insights into the processes that lead to it. 

We assume for simplicity a linear drag force of the form $-bV$ (see, e.g., \citealt{Chandra_1943}) and a gravitational force $G \Mblack M(R)/R^2$, where $\Mblack$ is the mass of the IMBH, $M(R)$ is the mass of the MW contained within the radius $R$, and $b$ is the drag force constant, with dimensions equal to a mass over a time. Note that a linear drag force is appropriate when the IMBH is moving through a non-turbulent medium at relatively low (radial) speed. In addition, we assume a MW with a constant mass density $\rho$. We let $t=0$ mark the time at which the IMBH entered the MW, and assume for simplicity $V(t=0) = 0$. We also let $R(t=0) = R_0$, and find a solution of the initial value problem of the form:
\begin{equation}
R(t) = \frac{R_0}{2\Phi} \, \left[ (1+\Phi)\, e^{-\frac{b}{2M}(1-\Phi)t} -(1-\Phi)\,e^{-\frac{b}{2M}(1+\Phi)t} \right] \, ,
\label{eq:R_physical}
\end{equation}
where:
\begin{equation} 
\Phi = \sqrt{1 - \frac{16 \pi G \rho M^2}{3 b^2}}
\label{eq:phi_eq}
\end{equation}
is a dimensionless parameter that controls the rate at which the radius decreases.

\begin{figure*}
	\includegraphics[width=2\columnwidth]{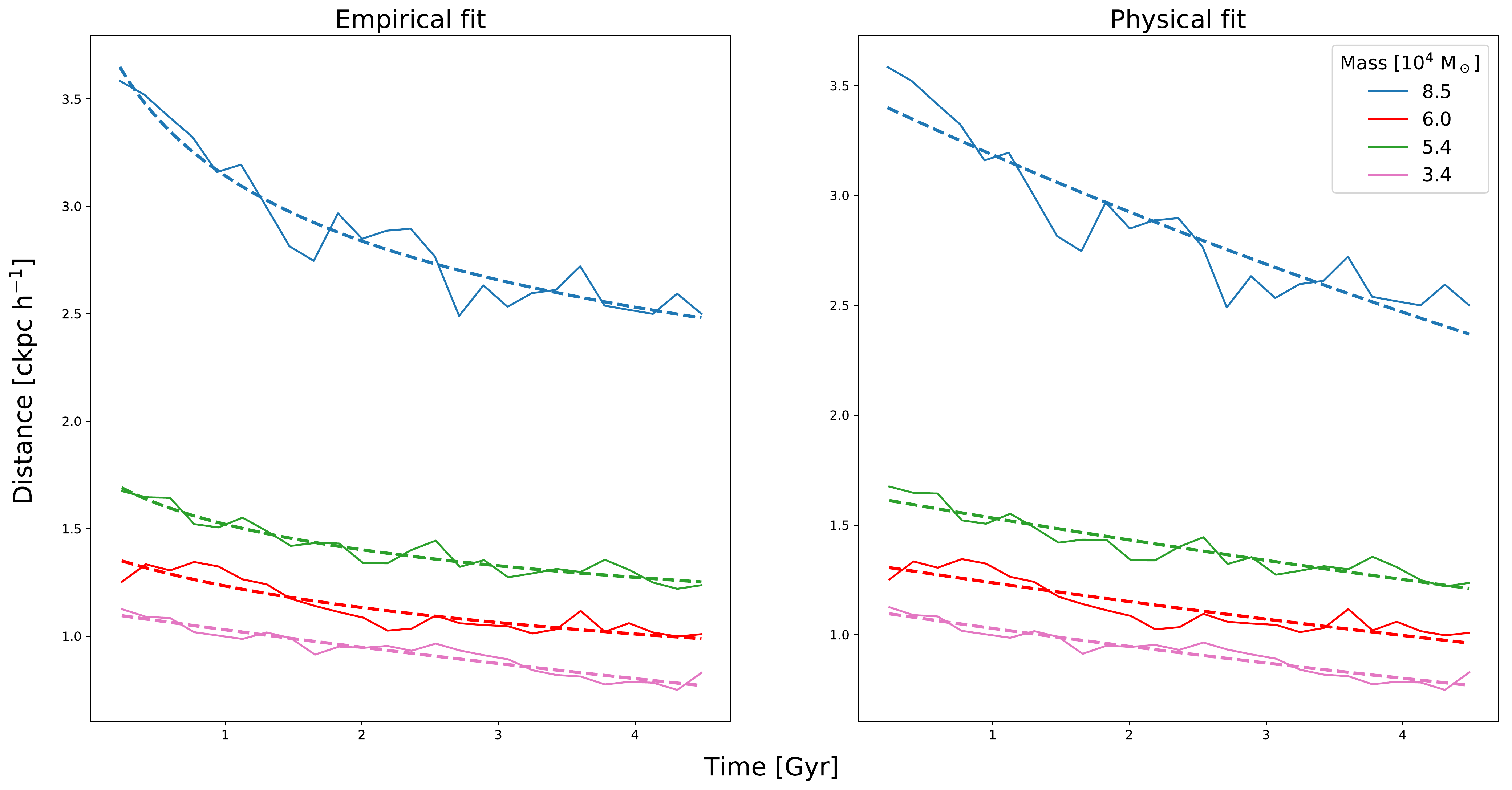}
    \caption{Radial distance vs. time for IMBHs in Set 1 (our smallest set, with the highest time resolution data). Fits from the empirical (Eq. \ref{eq:R_empirical}) and physical (Eq. \ref{eq:R_physical}) models are shown as dashed lines. Visually, both models are good fits for the data.}
    \label{fig:4 IMBHs with fit}
\end{figure*}

\begin{figure}
	\includegraphics[width=\columnwidth]{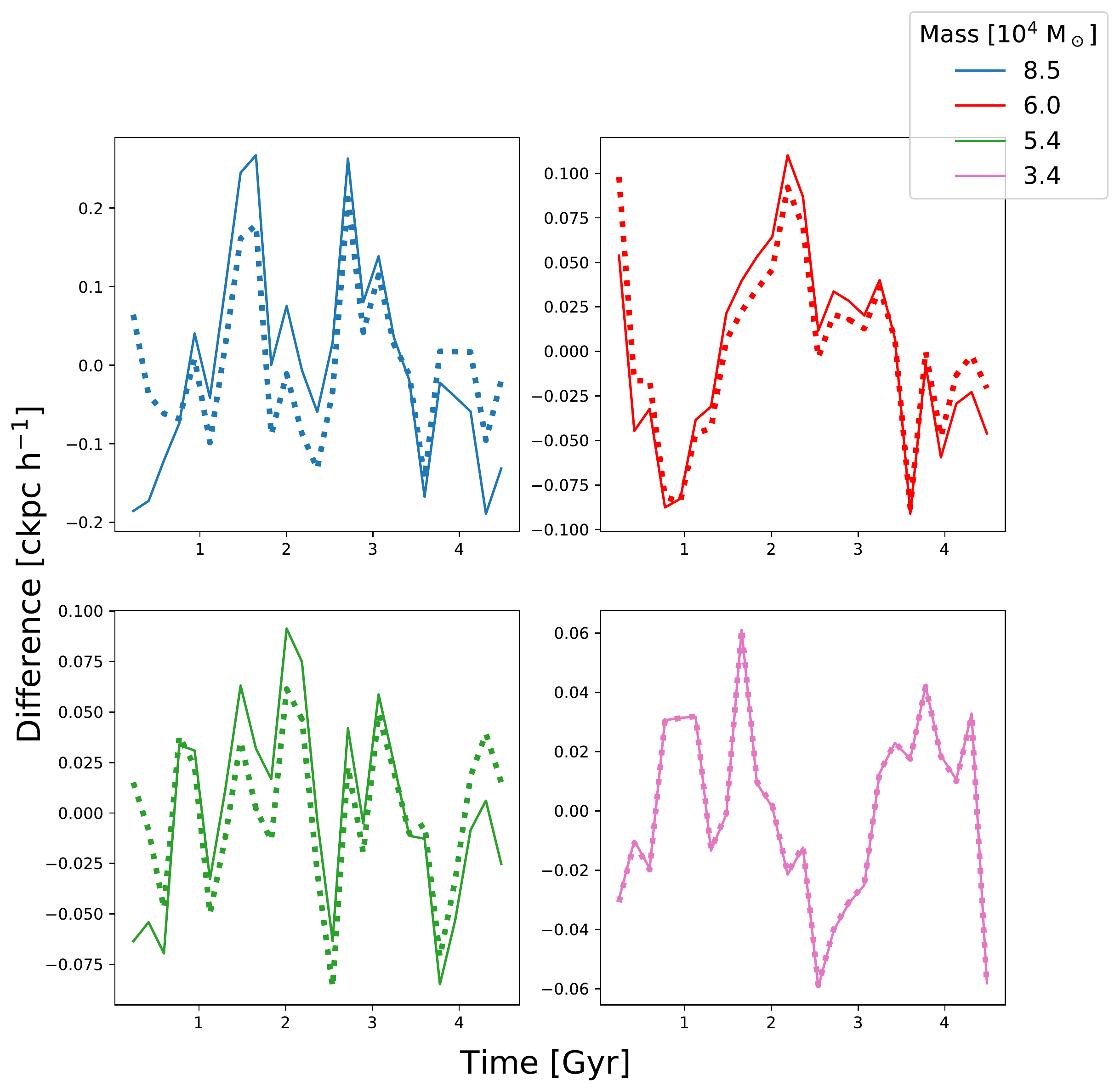}
    \caption{Difference between the empirical (dotted line) and physical (solid line) best fit curves and the simulation data for each IMBH in Set 1. The small residuals indicate that our physical and empirical models fit the data well.}
    \label{fig:Empirical and physical difference}
\end{figure}

Next, we found best fit curves of the forms in Eqs. \ref{eq:R_empirical} and \ref{eq:R_physical} for the radial distance vs. time data of each IMBH in Set 1. Here, $t$ represents the time since the dwarf galaxy hosting these four IMBHs merged with the MW (at snapshot 71, or $\sim 9.2 \, \rm Gyr$). The results are shown in Fig. \ref{fig:4 IMBHs with fit}, with the fit parameters for the physical model listed in Table \ref{tab:fit parameters}. Visually, both the empirical and physical models are successful in reproducing the decreasing distance of the IMBHs with increasing time. We also show in Fig. \ref{fig:Empirical and physical difference} the residuals between the physical and empirical models and the simulation data. These small differences support the conclusion that our models are good fits for the data.

We note that the values of the parameters $b/2M$ are consistent with each other, and the parameter of linear drag, $b$, is proportional to the mass of the perturber, $M$ \citep{BT_1987}. Interestingly, the values of the initial distances, $R_0$ are similar to each other, except for the heaviest IMBH which is approximately the double. Similar initial distances are expected, because all the IMBHs are originating from the same dwarf galaxy, which merges with the MW. We investigated this issue further, and found that the dwarf galaxy merged with the MW at snapshot 71, at which point the heaviest IMBH becomes classified as a particle belonging to the MW. However, the three lighter IMBHs become part of the MW only at the following snapshot 72. We hypothesize that some complex encounter dynamics may have influenced the association of the heaviest IMBH earlier than the others, so that its initial radius is also larger.

The physical model described by Eq. \ref{eq:R_physical} allowed us to quantify some relevant physical parameters. First, we used Eq. \ref{eq:phi_eq} and Table \ref{tab:fit parameters} to calculate $\rho$ from the fit parameters for each IMBH in Set 1. We found a mean density of $\sim 2.21 \times 10^{-4} \, \rm \Msun \, pc^{-3}$, with a standard deviation of $\sim 2.23 \times 10^{-5} \, \rm \Msun \, pc^{-3}$. For comparison, we calculated the average density of the MW analog within the half-mass radius, in each snapshot from 71 to 99. The mean density over these snapshots was $5.05 \times 10^{-5} \, \rm \Msun \, pc^{-3}$. Note that the half-mass radius of the MW is $\sim 38.4 \dist$, much larger than the radial distances of the IMBHs in Set 4, so it makes sense that the average density calculated using this half-mass radius would be lower than the estimate from our physical model. Therefore we conclude that our model provides a reasonable density estimate.

We also estimated the time needed for an IMBH to reach a radial distance of $1 \dist$, using Eq. \ref{eq:R_physical} and the mean values of $R_0$, $\Phi$, and $b/2M$ from Table \ref{tab:fit parameters}, finding a sinking timescale of $\sim 8.3 \, \rm Gyr$.

We note that in our analysis we focused on IMBHs with $v<0$, i.e. on sinking particles. Of course, as the drag force would work to oppose the velocity independently of the direction of the velocity vector, our model would apply also for particles with $v>0$. In fact, for IMBHs with positive radial velocities, their magnitudes are lower at smaller galactocentric distances.
In Fig. \ref{fig:Sinking velocity vs. distance fit} we limit our analysis to sinking IMBHs because: (i) we are mostly interested in particles that, in the long term, are approaching the center, because of the implications for the possibility of observing them, and (ii) they represent the majority of IMBHs studied.

The study of the effect of dynamical friction in the sinking (or not) of central black holes in merging galaxies is, of course, not new. Early works (e.g., \citealt{Governato_1994, Mayer_2007}) already showed the importance of dynamical friction from gas which leads central SMBHs in merging galaxies to approach each other, over sinking timescales of several Gyrs. More recent works have stressed the challenges in approaching distances shorter than $\sim 100$ pc in minor mergers (see, e.g., \citealt{Dosopoulou_2017}), the dependence on the mass ratio of the merging galaxies (see, e.g., \citealt{Tremmel_2018}), and the importance of the stellar component in the dynamical process, which can be dominant for IMBHs especially at high-$z$ (see, e.g., \citealt{Pfister_2019}).

\begin{table*}
\centering
\caption{Parameters for the physical model fit to the radial distance vs. time data for IMBHs in Set 1. Rounded to three decimals.}
\label{tab:fit parameters}
\begin{tabular}{cccc}
\hline
\textbf{Black hole mass} {[}$10^4 \Msun${]} & $\boldsymbol{R_0}$ {[}$\mathrm{ckpc \, h^{-1}}${]} & $\boldsymbol \Phi$ & $\boldsymbol{b/2M}$ {[}$\mathrm{Gyr}^{-1}${]} \\ \hline
$8.5$ & $3.461$ & $0.997$ & $27.543$ \\ 
$6.0$ & $1.327$ & $0.997$ & $27.066$ \\ 
$5.4$ & $1.636$ & $0.998$ & $27.114$ \\ 
$3.4$ & $1.117$ & $0.997$ & $27.043$ \\ \hline
\end{tabular}
\end{table*}

\section{Discussion and Conclusions} \label{sec:conclusions}

In this study we used the TNG50 cosmological simulation from the IllustrisTNG project to study the kinematics and dynamics of putative IMBHs wandering in the MW, as a result of their capture during mergers with dwarf galaxies. Some of the most relevant results are the following:
\begin{itemize}
    \item A fraction $\sim 86.6 \%$ of captured IMBHs in the MW are sinking;
    \item The highest density of IMBHs occurs in the central $1 \dist$ region of the galaxy, with a number density $\sim 1.6$ times higher than in the central $2 \dist$ region;
    \item The 3D velocities of IMBHs relative the MW center are $\gtrsim 2$ times larger than those relative to the local gas. Typical velocities with respect to the local gas are $\sim 88 \, \mathrm{km \, s^{-1}}$;
    \item Sinking velocities of IMBHs decline with decreasing radial distance, with a circularization of the orbits possibly due to drag effects.
\end{itemize}

These results will influence future studies of IMBHs wandering not only in the MW, but also in other galaxies. By locating the volume of the Galaxy containing the highest density of IMBHs, we provide guidance for observational efforts to detect these objects. Several studies have started to investigate the putative population of massive and supermassive wandering black holes in galaxies (see, e.g., \citealt{Bellovary_2010, Ricarte_2021a, Ricarte_2021b}). In particular, \cite{Bellovary_2010} predict that the MW should host 5-15 IMBHs within its halo, while more recent studies \citep{Ricarte_2021a} provide higher estimates.

Our results concerning velocity distributions relative to the galactic center and surrounding gas allow calculations of Bondi \citep{Bondi_1952} or Bondi-like accretion rates for IMBHs. Due to the low-density, non-central environments that should host these IMBHs, the appropriate mode to describe their accretion is an advection dominated accretion flow, or ADAF \citep{Narayan_1994, Narayan_1995, Abramowicz_1995, Narayan_2008, Yuan_Narayan_2014}, possibly modified to take into account the effects of convection and outflows \citep{Proga_2003, Igumenshchev_2003}. Note that already \cite{Bellovary_2010} predicted that off-nuclear black holes could be detected as ultra luminous X-ray sources.

Finally, the spatial and velocity distributions of these objects can inform predictions of gravitational wave events generated by merging IMBHs in the MW. It is possible that the current and future gravitational wave detectors such as LIGO, Virgo, KAGRA, and the Laser Interferometer Space Antenna (LISA, \citealt{LISA_2017}) could detect these merging events in the MW or nearby major galaxies (see, e.g., \citealt{Pacucci_2020}). For example, \cite{Fragione_2018} predict that up to $\sim 1000$ IMBHs, with masses in the range $10^3-10^4 \Msun$ in their model, could be wandering within $\sim 1 \, \mathrm{kpc}$ of the Galactic center, generating signals detectable in the gravitational waves domain. 

Of course, our findings come with several caveats. While we study over 2,000 IMBHs, several hundred of which came from the same dwarf galaxy, we are not suggesting that the MW, or dwarfs merging with the MW, contain this many actual IMBHs. Rather, our study provides a statistical sampling of the location and velocity distribution of IMBHs in the MW, without offering any insights on their actual number.

As explained in \S \ref{sec:proxies}, the IMBHs used are proxies, i.e. stellar objects in the correct mass range. Their gravitational effect on the environment should be the same, as noted also by, e.g., \cite{Mistani_2016}, where the authors used the same technique for globular clusters instead of IMBHs. Nevertheless, this is a point worth remembering when considering our results. One effect of using star clusters as proxies, also discussed in \S \ref{sec:proxies}, is that they only cover a fraction of the typical mass range of IMBHs. 
Additionally, cluster objects are subject to some mass loss through evaporation. In our simulations, IMBH proxies have a median mass loss of $\sim 3.6\%$ over the time frame investigated (snapshots 81-99), which is too small to affect our conclusions. 
Finally, we note that our kinematic and dynamical analyses are constrained by a time resolution of $\sim 0.14 \, \rm Gyr$ in the main simulation, and $\sim 0.004 \, \rm Gyr$ in sub-box 2.

In summary, this study is an important first step in investigating the properties of captured IMBHs in the MW. Despite large unknowns in their properties and regarding their very existence in the MW, efforts to detect them are underway, and are crucial to understand the last unknown black hole population in our Galaxy.

\section*{Acknowledgements}
We thank the anonymous referee for constructive comments on the manuscript. The authors acknowledge support and fruitful discussions with Annalisa Pillepich, Dylan Nelson and Ramesh Narayan. E.W. acknowledges undergraduate research support provided in part by the Harvard College Program for Research in Science and Engineering (PRISE).
F.P. acknowledges support from a Clay Fellowship administered by the Smithsonian Astrophysical Observatory. 
S.B. is supported by the UK Research and Innovation (UKRI) Future Leaders Fellowship [grant number MR/V023381/1].
This work was partly performed at the Aspen Center for Physics, which is supported by National Science Foundation grant PHY-1607611. The participation of F.P. at the Aspen Center for Physics was supported by the Simons Foundation.
This work was also supported by the Black Hole Initiative at Harvard University, which is funded by grants from the John Templeton Foundation and the Gordon and Betty Moore Foundation. The IllustrisTNG simulations were undertaken with compute time awarded by the Gauss Centre for Supercomputing (GCS) under GCS Large-Scale Projects GCS-ILLU and GCS-DWAR on the GCS share of the supercomputer Hazel Hen at the High Performance Computing Center Stuttgart (HLRS), as well as on the machines of the Max Planck Computing and Data Facility (MPCDF) in Garching, Germany.

\section*{Data Availability}
The data underlying this article are available on the website of the \href{https://www.tng-project.org/}{\color{cyan}IllustrisTNG project}, as described in \cite{Nelson_2019_Illustris}. The codes used to analyze the data will be shared on reasonable request to the corresponding author.



\bibliographystyle{mnras}
\bibliography{ms} 





\bsp	
\label{lastpage}
\end{document}